# MICROSTRUCTURAL CHARACTERISTICS OF REACTION-BONDED B$_4$C/SiC COMPOSITE

Tianshi Wang[1], Chaoying Ni[1*], Prashant Karandikar[1,2]

[1]Department of Materials Science and Engineering, University of Delaware, Newark, DE 19716, USA
[2]M-Cubed Technologies, Inc., 1 Tralee Industrial park, Newark, DE 19711, USA

## Abstract

A detailed microstructural investigation was performed to understand structural characteristics of a reaction-bonded B$_4$C/SiC ceramic composite. The state-of-the-art focused ion beam & scanning electron microscopy (FIB/SEM) and transmission electron microscopy (TEM) revealed that the as-fabricated product consisted of core-rim structures with α-SiC and B$_4$C cores surrounded by β-SiC and B$_4$C, respectively. In addition, plate-like β-SiC was detected within the B$_4$C rim. A phase formation mechanism was proposed and the analytical elucidation is anticipated to shed light on potential fabrication optimization and the property improvement of ceramic composites.

* Corresponding author. Tel 1 302 831 3569; Fax 1 302 831 4545

## Introduction

Reaction-bonded B$_4$C/SiC ceramic composites were fabricated by infiltrating molten Si into a preheated preform consisting of B$_4$C, α-SiC and C to form a composite of near theoretical density [1]. The 10-15 vol% residual Si, together with other newly formed phases, bond the preexisting B$_4$C and α-SiC to produce a cohesive solid of advanced properties, including light weight (~2.8 g/cm$^3$), high thermal stability, and corrosion resistance, in addition to high mechanical properties such as Young's modulus (~420 GPa) [2,3], resulting in applications for armor, thermal management, wear, and precision equipment.

Due to different processing conditions and parameters, varied phase evolution and microstructures of B$_4$C/SiC ceramic composites were reported. Ness studied reaction-bonded SiC and found that the newly formed β-SiC grew epitaxially on original α-SiC [4]. To explain the structural evolution, Ness proposed a dissolution-precipitation mechanism for newly formed β-SiC. To understand B$_4$C-Si reaction, Teller sintered B-rich corner in the phase diagram of B-C-Si system to determine equilibrium phases and developed a theory for the formation of Si dissolved B$_4$C expressed as B$_{12}$(Si,B,C)$_3$ [5]. Hayun studied the microstructure of reaction-bonded B$_4$C (RBBC) and observed plate-like β-SiC as well as a core-rim microstructure consisting of B$_4$C grains surrounded by secondary B$_{12}$(Si,B,C)$_3$ which he assumed was formed by a dissolution-precipitation mechanism [6,7]. In a study by Karandikar, however, jagged B$_4$C microstructures were observed in RBBC which he suggested that a diffusion mechanism was more plausible to explain the rim formation of Si containing B$_4$C [8]. Due to the complexity of high temperature chemical reactions, phase transformation, and the resulting phase structures, there are still significant gaps to bridge between the macrostructural characterization, phase evolution, and microscopic mechanisms.

In this study, a reaction-bonded $B_4C/SiC$ composite was characterized by utilizing the cutting-edge FIB/SEM and TEM to precisely determine the composite constituents. An assessment was made based on the reaction bonding thermodynamics to elucidate in detail the diffusion, interactions, and phase transformations.

## Experimental Procedure

The reaction-bonded $B_4C/SiC$ composites were fabricated by infiltrating a preform of $B_4C$, α-SiC, residual C and 20-25 vol.% pores with molten Si at about 1500°C. Both the C and the porosity were left from preheating an organic binder used to form the green body of $B_4C$ and α-SiC [9,10]. The reactions between the infiltrating liquid Si and the preform led to a dense product with desired near-net shape and dimension.

The microstructure of the samples was studied using a FIB/SEM (Zeiss Auriga 60) with an X-ray energy-dispersive spectrometer (EDS, Oxford Instruments X-Max 80) and a transmission electron microscope (TEM, JOEL JEM-2010F). SEM samples were prepared using a standard metallographic procedure that included a final polish with 1 μm diamond suspension (Buehler MetaDi). An X-ray diffractometer (Bruker D8) was used to determine phases with a position sensitive detector (LynxEye).

TEM samples were prepared using the FIB/SEM. Two trapezoids along each side of selected areas were first milled to obtain a lamella of about 1 μm thick, and the lamella was then further tilted to mill the cross-section. The lamella was thinned to a thickness of less than 100 nm by Ga ion beam of adequate energy. After that, the lamella was mounted to an OmniProbe (Oxford Instruments) by Pt deposition and then transferred to a copper grid.

## Results and discussion

Phase Structures

Fig 1(a) is a typical SEM image of the composite microstructure. Four phases are identified based on grain morphology and contrast, together with the EDS mapping in Fig. 1(c)-(e), and these phases are $B_4C$ (dark regions), SiC (grey grains), residual Si (bright regions), and the plate-like SiC which usually appears inside $B_4C$. The image contrast here is primarily from the average atomic number variation of phase constituents. No pores were detected by SEM as anticipated based on the process parameters.

Further analysis with an in-lens detector, which offers better topographic contrast, revealed the core-rim structures associated with most of $B_4C$ grains and also with a considerable number of SiC grains, as shown in Fig. 1(b). Fig. 2 shows the XRD pattern of the sample and the peaks are indexed as α-SiC, β-SiC, $B_4C$, Si, and $B_{12}(Si,B,C)_3$.

To identify the formation mechanisms of the β-SiC, rim structures surrounding the original α-SiC and B₄C core grains, we proceeded with additional SEM and TEM investigations.

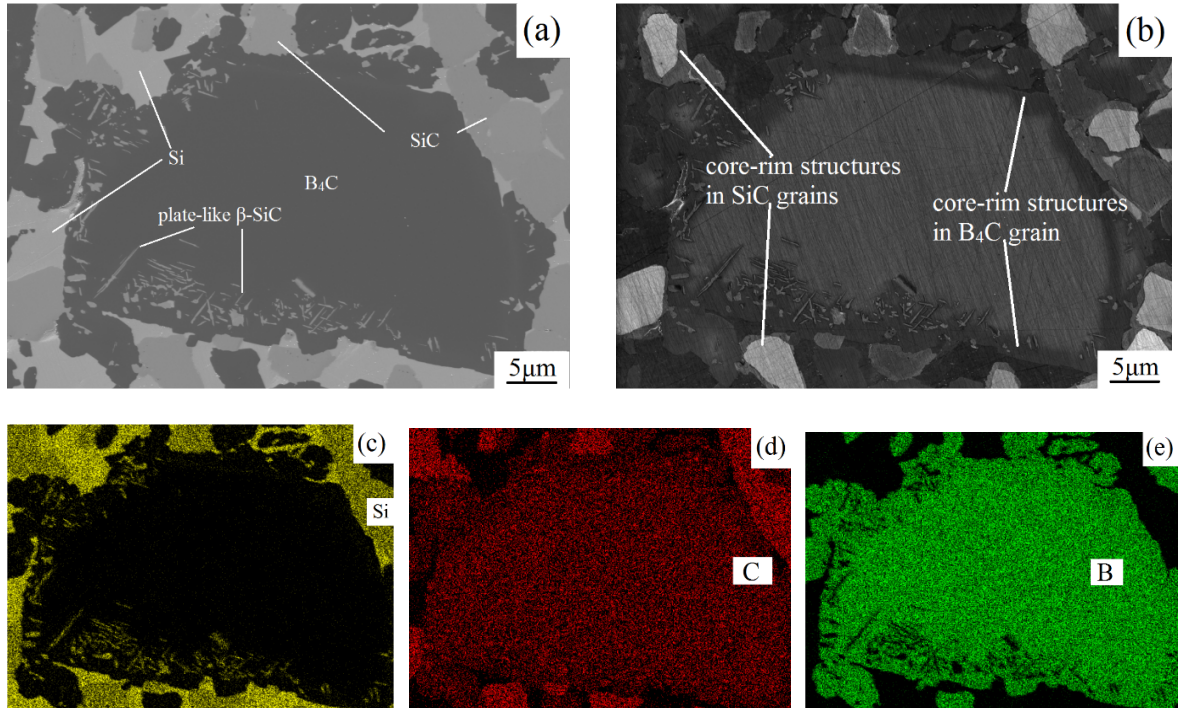

Fig. 1. SEM micrographs and EDS maps of the core-rim structures of B₄C and SiC. Image (a) and (b) were captured by a chamber secondary electron detector (SESI) and an in-lens detector respectively. Images (c-e) were EDS elemental maps of Si, C, and B respectively. The in-lens image (b) shows additional structural contrast within the grains, while the SESI image (a) and EDS mapping (c-e) do not.

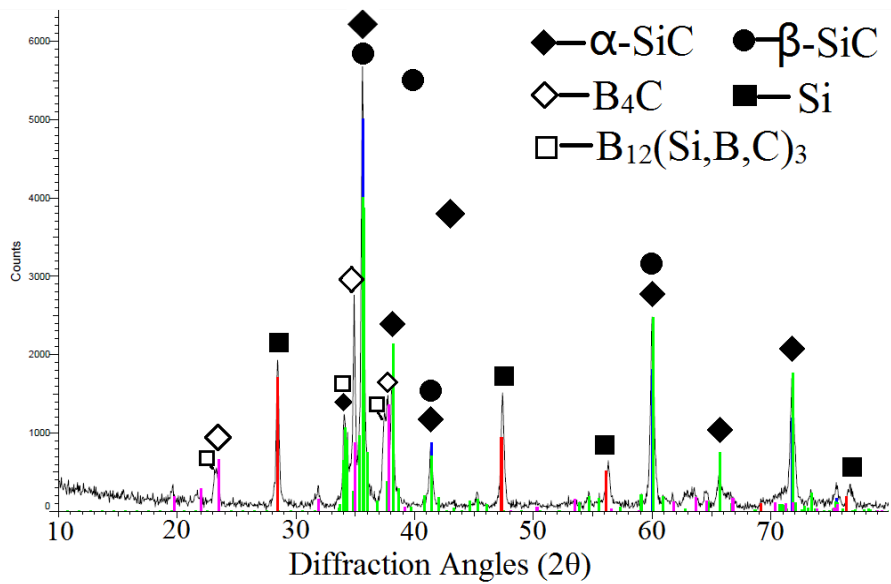

Fig. 2. X-ray diffraction pattern of the reaction-bonded SiC and B$_4$C composite. Five phases are identified.

Core-rim Structure of SiC

A core-rim structure of SiC is shown in Fig. 3(a). The rectangular inset is an image from a section polished by FIB while the rest of the image field of view is from a sample mechanically polished in a conventional way. It is noticed that the 4 original α-SiC grains are interconnected by the rim structure to form a cluster of SiC grains. The secondary electron image collected from SESI and the EDS mapping indicate negligible contrast difference between the newly formed β-SiC and the original α-SiC (Fig. 3b-c), suggesting a minimal compositional difference between the two types of SiC grains.

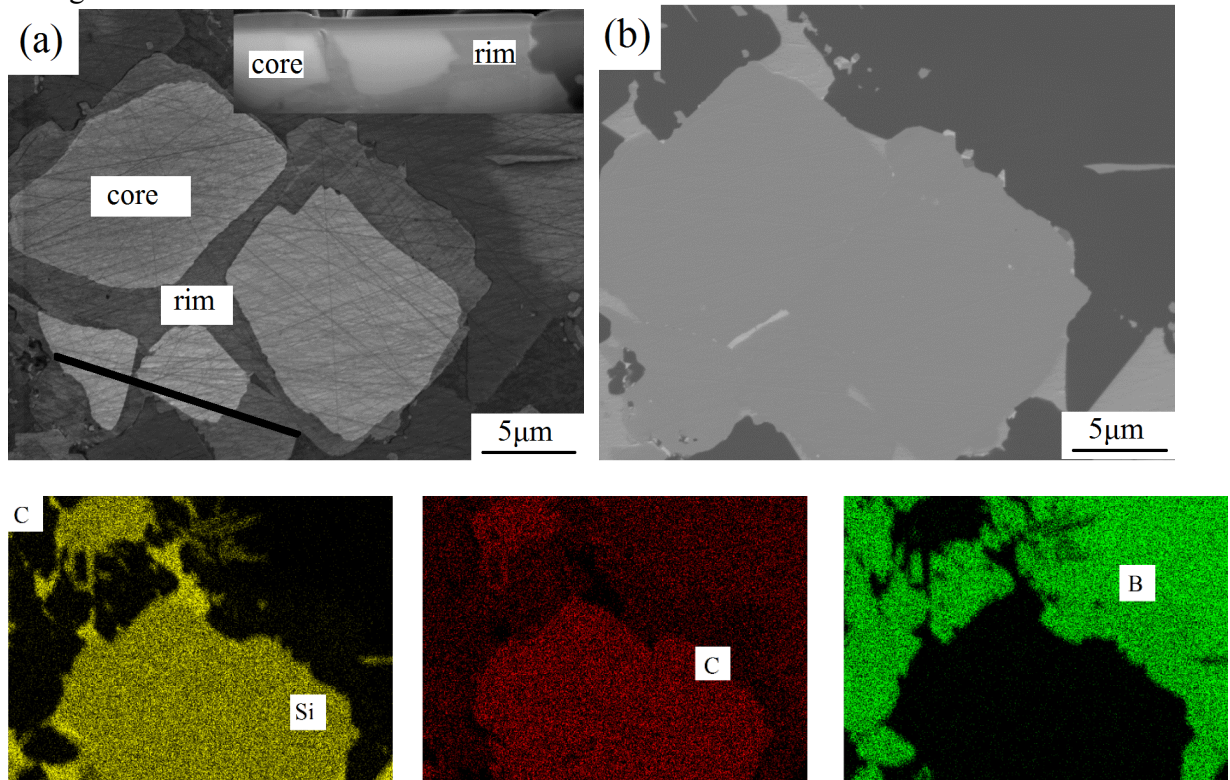

Fig. 3. (a) An in-lens image of the core-rim structure of SiC with an inset of the FIB polished cross-section; (b) chamber detector (SESI) image of the same area; (c) EDS mapping for C, Si and B, respectively. Both Si and C distribute evenly inside SiC grains, the original as well as the newly formed.

To further understand the SiC core-rim structure, we performed TEM. The upper-left area in Fig. 4 is in the SiC core while the lower-right area is in the SiC rim. TEM suggests that no obvious faults are detectable in the core area. However, the β-SiC rim is filled with faults and crystal defects, which also explains the topographical difference after the polishing that in turn contributes to the contrast of the in-lens image.

It is plausible that the β-SiC rim surrounding α-SiC grows based on the α-SiC core through a solidification process of the molten Si[C]. At the beginning, some residual carbon would dissolve in the molten Si and generate a great amount of heat as dictated by the reaction thermodynamics. The heat further increases the dissolution of C in Si; therefore, the local temperature and C concentration in the reaction region are high. With a high diffusion coefficient at such high temperature, C should spontaneously diffuse to the vicinity of α-SiC where the temperature and C solubility are lower and the resulting supersaturated liquid Si[C] therefore heterogeneously crystalizes to form β-SiC attached to the original α-SiC grains, a preferred process due to relatively lower energy in contrast to the homogeneous nucleation in liquid phase. The newly formed β-SiC grows fast under a much lower temperature than that of the original α-SiC [11]. Such a mechanism of fast growth promotes a large number of faults inside the newly formed SiC grains.

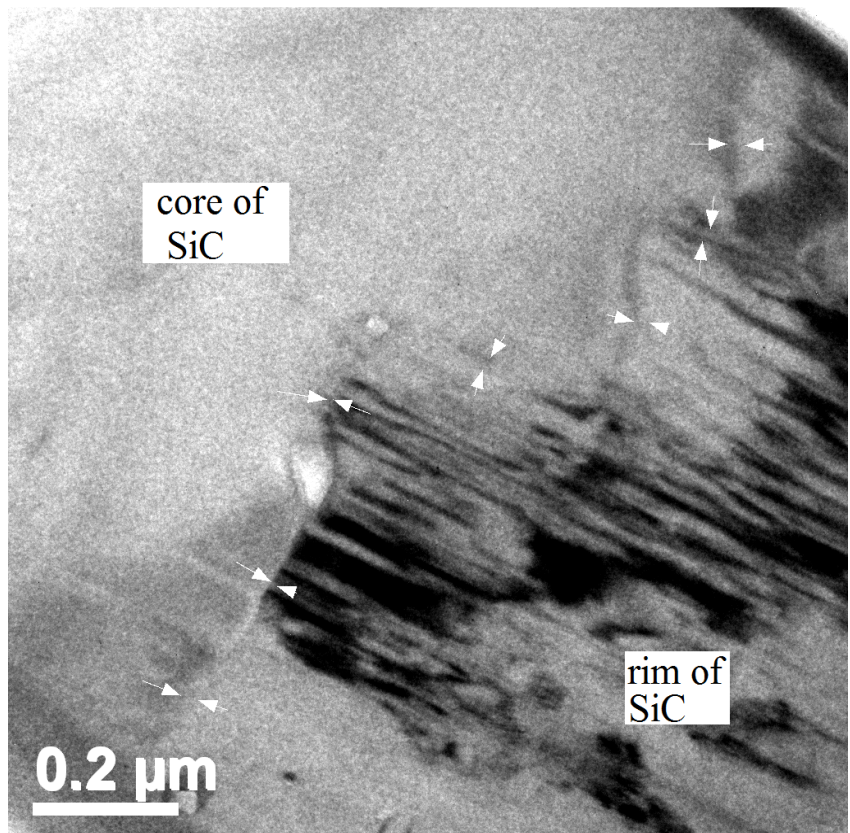

Fig. 4. TEM image of core-rim structure of SiC. Upper-left is the original α-SiC core area appearing free of crystal defects while the lower-right is the newly formed β-SiC, in which severe twinning and lattice distortion are visible. Arrows indicate the grain boundary of the two distinct areas.

Plate-like β-SiC

The plate-like β-SiC commonly appears in the rim of $B_4C$ as shown in Fig. 1(a-c). Both of the plate-like and the occasional needle-like morphologies were observed by SEM after FIB milling of the β-SiC cross sections. A bright field TEM image of two β-SiC grains inside $B_4C$ are shown in Fig. 5(a). Defects, especially stacking faults, are easily noticeable in the β-SiC. Fig. 5(b) is a bright field image of a grain boundary between β-SiC and $B_4C$.

Based on the SEM and TME observation and analysis, the formation mechanism of β-SiC is proposed as follows:

As the infiltrating Si meets the residual C in the green body, the molten Si dissolves C to form liquid Si[C], which is a highly exothermic process. Since solid SiC is more stable than $B_4C$ at around 1600°C according to Si-C-B diagram [12] and the original $B_4C$, α-SiC and their surroundings are relatively away from the heat source of residual C, the β-SiC heterogeneously crystalizes on or in the vicinity of α-SiC and $B_4C$ and grows into liquid Si[C] especially during the cooling process. While the formation of β-SiC surrounding $B_4C$ occurs at a relatively higher temperature and the limiting factor of β-SiC growth is C concentration, the β-SiC grains tend to grow toward the C supersaturated liquid as plates with significant growth defects such as twinning and stacking faults. Defects are also observed in the β-SiC rim surrounding α-SiC presumably due to similar fast growth and thermal fluctuation.

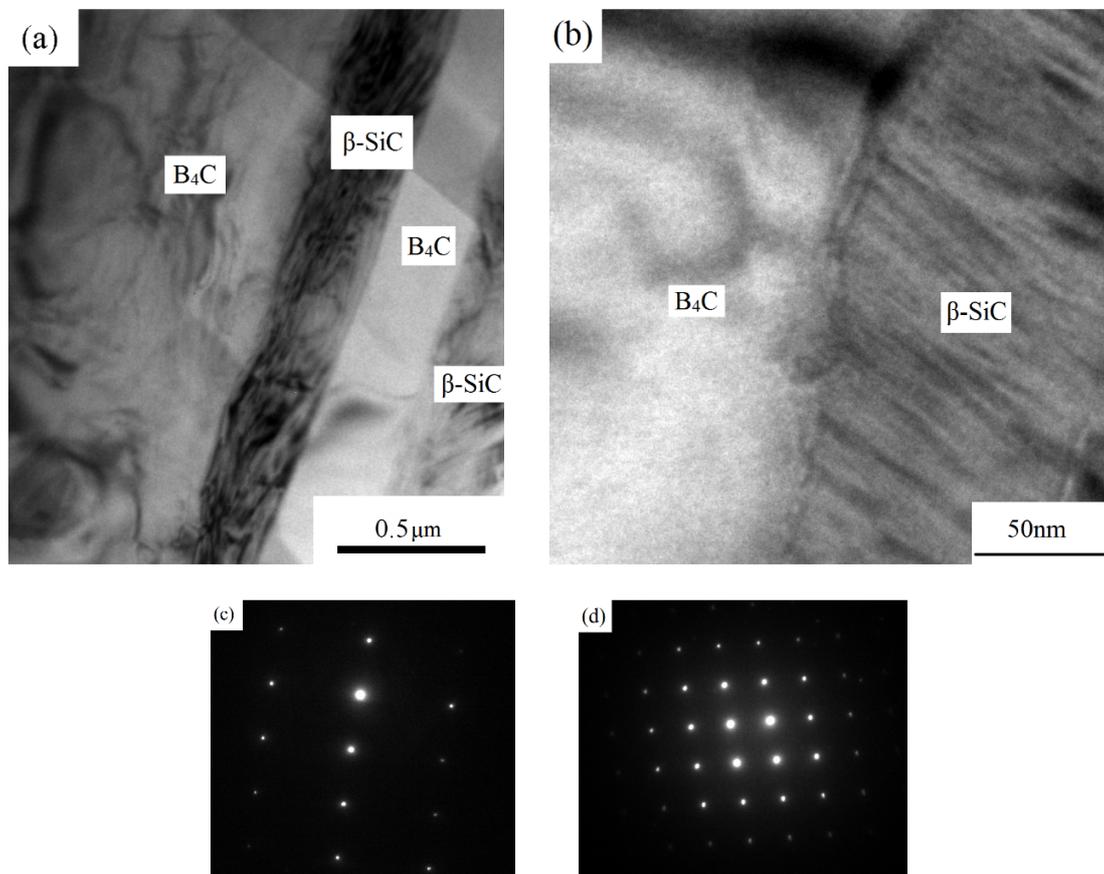

Fig. 5. (a) TEM bright field image of β-SiC inside $B_4C$; (b) The boundary between β-SiC and $B_4C$; (c) Diffraction pattern of β-SiC; (d) Diffraction pattern of $B_4C$.

Core-rim Structure of $B_4C$

Boron carbide rims were frequently observed around $B_4C$ with relatively lower brightness in the secondary electron contrast. In addition, the β-SiC plates usually exist inside the rim and some of them also appear in the original $B_4C$.

An EDS line-scan along the black line in Fig. 6(a) shows that in this $B_4C$ rim, Si maintains a very low concentration of around 2 at.% in the rim to form $B_{12}(Si,B,C)_3$ as also identified by XRD. It was suggested that the rim phase was formed by Si diffusion from liquid [8]. While the Si inward diffusion in $B_4C$ indeed appears to be a plausible mechanism, additional investigation is necessary to fully exclude the possibility of $B_{12}(Si,B,C)_3$ re-precipitation following the $B_4C$ dissolution as was suggested in literature [6,7]. It is believed also possible that the $B_4C$ rim structure is formed by a process of simultaneous Si inward diffusion and limited $B_{12}(Si,B,C)_3$ growth within and towards the liquid Si[B,C] formed through an earlier dissolution of C and $B_4C$ into the liquid Si. The plate-like β-SiC contained in the $B_{12}(Si,B,C)_3$ rim and the original $B_4C$ could be secondary precipitates formed during the cooling process and these plates also appear to preferentially distribute along specific crystal planes of the $B_4C$ and $B_{12}(Si,B,C)_3$ matrixes.

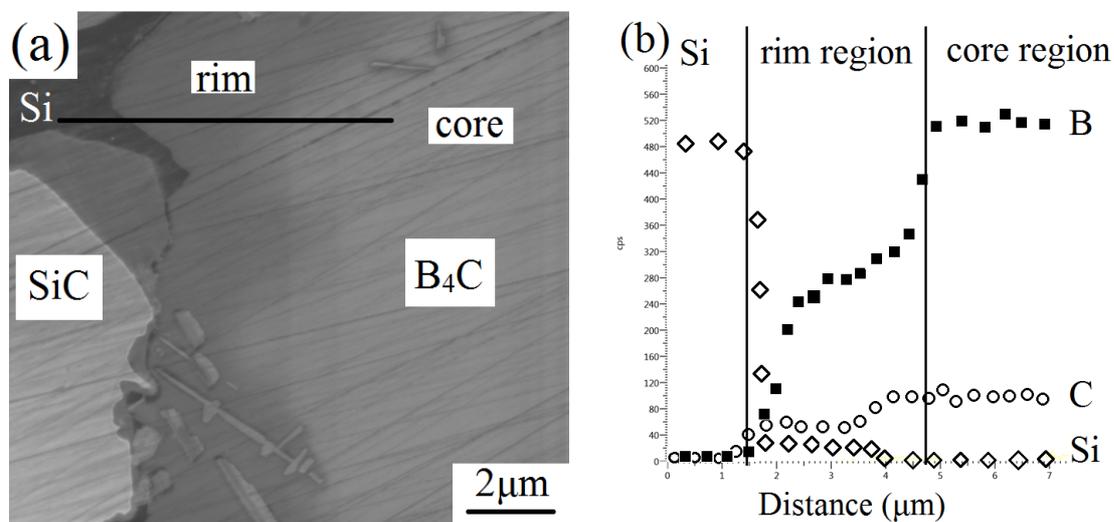

Fig. 6. Elemental profiles of EDS line-scan along the reference line (a) and the data plot (b).

## Conclusion

The microstructural characteristics of a reaction bonded $B_4C$/SiC CMC were investigated. Apart from $B_4C$, α-SiC, and Si, we observed plate-like β-SiC and core-rim structures surrounding both SiC and $B_4C$ grains. With TEM, we discovered significant faults and defects inside β-SiC plates and also in the β-SiC rim surrounding α-SiC. Based the morphology and thermodynamics, we propose that the rim of α-SiC is formed by precipitation of supersaturated Si[C] and the $B_{12}(Si,B,C)_3$ rim surrounding original $B_4C$ is generated by means of a dominant Si diffusion. The β-SiC plates in $B_4C$ and $B_{12}(Si,B,C)_3$ appear to precipitate during cooling and to distribute along certain crystal planes. The results are anticipated to help make potential improvements of the fabrication process and composite properties.

## Acknowledgement

This work is supported by the II-VI Foundation. Instrumentation and staff supports from the W. M. Keck Center for Advanced Microscopy and Microanalysis at the University of Delaware are greatly appreciated.

# References

[1]   P. Karandikar et al., "Diamond-Reinforced Composite Materials and Articles, and methods for making same," US 8,474,362 (2013) 1-20.

[2]   S.M. Salamone et al., "Effect of SiC:$B_4$C ratio on properties of Si-Cu/SiC/$B_4$C composites," Mech. Prop. Perform. Eng. Ceram. Compos. IX. (2014) 83–90.

[3]   S.M. Salamone et al., "Macroscopic assessment of high pressure failure of $B_4$C and $B_4$C/SiC composites," Adv. Ceram. Aror IX Ceram. Eng. (2013) 25–30.

[4]   J.N. Ness, T.F. Page, "Microstructural Evolution in Reaction-bonded Silicon Carbide," J. Mat. Sci. 21 (1986) 1377–1397.

[5]   R.Telle, "STRUCTURE AND PROPERTIES OF SI - DOPED BORON CARBIDE," Phys. Chem. Carbides, Nitrides Borides, Springer Netherlands, 1990: pp. 249–267.

[6]   S. Hayun et al, "Rim region growth and its composition in reaction bonded boron carbide composites with core-rim structure," J. Phys. Conf. Ser. 176 (2009) 012009.

[7]   S. Hayun et al., "Microstructural evolution during the infiltration of boron carbide with molten silicon," J. Eur. Ceram. Soc. 30 (2010) 1007–1014.

[8]   P.G. Karandikar et al., "Microstructural Development and Phase Changes in Reaction Bonded Boron Carbide," Adv. Ceram. ARMOR VI Ceram. Eng. Sci. Proc. 31 (2010) 251–259.

[9]   M. Aghajanian, A. McCormick, "Composite Materials and Articles, and methods for making same," US 8,128,861 B1, 2012.

[10]  M. Aghajanian et al., "Boron carbide composite bodies, and methods for making same," US 6,862,970 B2, 2005.

[11]  A.S. BAKIN, "SiC HOMOEPITAXY AND HETEROEPITAXY," Int. J. High Speed Electron. Syst. 15 (2005) 747–780.

[12]  E. Gugel et al., "Investigations in the ternary system boron--carbon--silicon," Solid State Chem. Proc. 5th Mater. Res. Symp., National Bureau of Standards, 1972: pp. 505–513.